\documentstyle[aps]{revtex}

\begin{document}
\draft
\title{Spin-lattice models: inhomogeneity and diffusion}
\author{Han Zhu$^1$ and Jian-Yang Zhu$^{2,3\thanks{%
Author to whom correspondence should be addressed. Email address:
zhujy@bnu.edu.cn}}$}
\address{$^1$Department of Physics, Nanjing University, Nanjing, 210093, China\\
$^2$CCAST (World Laboratory), Box 8730, Beijing 100080, China\\
$^3$Department of Physics, Beijing Normal University, Beijing 100875, China}
\maketitle

\begin{abstract}
In spin-lattice models with order parameter conserved, we generalize the
idea of spin diffusion incorporating a variety of factors as possible
driving forces, including external field and temperature. The Kawasaki
dynamics in the Gaussian model and the one-dimensional Ising model are
studied as specific examples. In the obtained diffusion equation, the term
describing the diffusion induced by the inhomogeneity of the magnetization
itself is unaffected and is believed to vanish near the critical point.
Meanwhile the nonvanishing diffusion induced by the inhomogeneity of the
environment may be coupled to the spin configuration and weakened by thermal
noise. Interesting dynamic behavior is observed as a result of a competition
of internal and external inhomogeneities and time scales. Several
interesting examples are visualized, and the concept of local hysteresis is
proposed in this spin-conserved dynamics.
\end{abstract}


\section{Introduction}

\label{Section 1}

In the past decade, great interest has been aroused by the response of a
cooperative system to external perturbations, e.g. an external field varying
with time or a scanning temperature (see Ref. \cite
{dyna-tran,co-expe,xy,the-hys-fe,the-hys-cdw,the-hys-1,the-hys-2} and
references therein). In particular, significant effort has been made in the
dynamic phase transition and hysteresis of the Ising model subject to an
oscillating field\cite{dyna-tran}, using Monte Carlo simulation and
Mean-Field approximation based on Glauber's dynamics. Yet the dynamics with
conservation law\footnote{%
In the absence of driving force, the Ising model with spin conserved is
relatively well understood: Quenched from a high temperature disordered
state, some kind of order rises with the appearance of coarsening and
competing up-and-down domains, of which the length scale grows as a power
law with time\cite{quenched-1,quenched-2}.}, i.e. a model B in the
classification of Hohenberg and Halperin\cite{class}, has been much less
studied. Such dynamics may describe, for example, the phase separation of a
binary mixture subject to inhomogeneous temperature or gravity.

Generally speaking, in spin-lattice models with the order parameter
conserved, it is possible to describe a diffusion process with a simple
diffusion equation. In homogeneous environment, it is natural to guess that
the spin flow is proportional to the gradient of the spin, and the equation
will be 
\begin{equation}
\frac{\partial q\left( {\bf r},t\right) }{\partial t}={\cal D}\nabla
^2q\left( {\bf r},t\right) ,  \label{simple}
\end{equation}
where $q\left( {\bf r},t\right) $ denotes the local magnetization and ${\cal %
D}$ is a diffusion coefficient. This equation has been exactly obtained in
the Gaussian model and ${\cal D}$ is found to vanish at the critical point
(see Ref. \cite{spin-pair}, and the following section). It has also been
studied in other systems, including the Ising model\cite{Kawasaki}.

Diffusion is always a result of inhomogeneities. The diffusion process
characterized by Eq. (\ref{simple}) can be viewed as being caused by the
inhomogeneity of the magnetization itself, while diffusion driven by
external inhomogeneities remains to be an unsolved interesting problem. Many
studies on the dynamic response of the Ising model deal with a uniform field
varying with time\cite{dyna-tran}. However, such a field has no effect on
the Kawasaki dynamics\cite{Kawasaki} with spin conservation. So here is a
more difficult task, which should incorporate both spatial and temporal
inhomogeneities. In relevant studies several interesting problems have been
treated, e.g., the random-field spin-conserving Ising model\cite
{ran-fie-1,ran-fie-2,ran-fie-3,ran-fie-4} and, more recently, the domain
growth in a temporally constant field varying linearly in space\cite
{driven-ising-1,driven-ising-2}. In the present study, we narrow our scope
to the Gaussian model and the one-dimensional Ising model, and focus on the
modification of the above diffusion equation by the inhomogeneity of field
and temperature.

As is well known, the Ising model has wide and important applications in
various fields. Unfortunately, as we shall see below, the obtained diffusion
equation can not be solved. Meanwhile, it is analytically tractable in an
idealization of the Ising model, the kinetic Gaussian model, which may serve
as a starting point and a test field of new ideas.

In this article, the rigorous treatment of the Gaussian model will be
combined with the less quantitative discussions and Monte Carlo simulations
of the Ising model, along the same line, for a comparison. Section \ref
{Section 2} deals with the diffusion induced by inhomogeneous external
field, and in Sec. \ref{Section 3} we extend the scope to inhomogeneous
temperature. (Although here we treat these two problems separately, it is
possible to put them together.) Section \ref{Section4} is the summarization
with some discussions.

\section{Inhomogeneity of the external field}

\label{Section 2}

In this section, based on the master equation, we shall study diffusion
process in an inhomogeneous external field. Such a process is subject to two
factors acting together, the inhomogeneity of the field and that of the
magnetization itself. This competition may lead to interesting and rich
dynamic behavior, as we shall see below. Before we move to the specific
models, we briefly review the governing dynamic mechanism, spin-pair
redistribution mechanism, which is a natural generalization of Kawasaki's
spin-pair exchange mechanism\cite{Kawasaki}.

In the Ising model, spins can only take two opposite values, $+1$ and $-1$,
and spin-exchange is simply a natural choice; but it turns to be not as
capable when applied to other more complex systems. A natural generalization
is as follows: a pair of nearest neighbors selected at random, $\sigma _i$
and $\sigma _j$, may take any values, $\hat{\sigma}_i$ and $\hat{\sigma}_j$,
as long as their sum keeps conserved, and this is called spin-pair
redistribution (clearly in the Ising model it reduces to the spin-exchange).
We list several important equations in Appendix \ref{Appendix}, the details
of which can be found in Ref. \cite{spin-pair}.

\subsection{The Gaussian model}

\label{subsection 2.1}

First we will study the three-dimensional kinetic Gaussian model in an
inhomogeneous external field, which might also be varying with time (e.g., a
traveling wave). In lattice models, a traditional way is to assign to each
vertex, $\sigma _i$, a reduced field, $H_i\left( t\right) $\cite{dyna-tran}.
In the kinetic Gaussian model, 
\begin{equation}
-\beta {\cal H}=K\sum_{\left\langle i,j\right\rangle }\sigma _i\sigma _j+%
\frac 1{k_BT}\sum_iH_i\left( t\right) \sigma _i,  \label{G-Ham}
\end{equation}
where $K=J/k_BT$, and $J$ is the coupling strength. The spins take
continuous value from $-\infty $ to $+\infty $, and the probability of
finding a given spin between $\sigma _i$ to $\sigma _i+d\sigma _i$ is
assumed to be the Gaussian-type distribution $f\left( \sigma _i\right)
d\sigma _i=\sqrt{\frac b{2\pi }}\exp \left( -\frac b2\sigma _i^2\right)
d\sigma _i$, where $b$ is a distribution constant independent of
temperature. Thus, a summation over the spin values will be in fact an
integration, $\sum_\sigma \rightarrow \int_{-\infty }^\infty f\left( \sigma
\right) d\sigma $.

Now, we substitute the Hamiltonian (\ref{G-Ham}) to the evolving equation of
single spin, Eq. (\ref{ssr-q}). A linear equation can be obtained
(three-dimensional) 
\begin{eqnarray}
\frac d{dt}q_{ijk}\left( t\right) &=&\frac 1{2\left( b+K\right) }\left\{
b\left[ \left( q_{i+1,j,k}-2q_{ijk}+q_{i-1,j,k}\right) \right. \right. 
\nonumber \\
&&\left. +\left( q_{i,j+1,k}-2q_{ijk}+q_{i,j-1,k}\right) +\left(
q_{i,j,k+1}-2q_{ijk}+q_{i,j,k-1}\right) \right]  \nonumber \\
&&+K\left[ 2\left( 2q_{i-1,j,k}-q_{i-1,j+1,k}-q_{i-1,j-1,k}\right) +\left(
2q_{i-1,j,k}-q_{ijk}-q_{i-2,j,k}\right) \right.  \nonumber \\
&&+2\left( 2q_{i+1,j,k}-q_{i+1,j+1,k}-q_{i+1,j-1,k}\right) +\left(
2q_{i+1,j,k}-q_{ijk}-q_{i+2,j,k}\right)  \nonumber \\
&&+2\left( 2q_{i,j-1,k}-q_{i,j-1,k+1}-q_{i,j-1,k-1}\right) +\left(
2q_{i,j-1,k}-q_{ijk}-q_{i,j-2,k}\right)  \nonumber \\
&&+2\left( 2q_{i,j+1,k}-q_{i,j+1,k+1}-q_{i,j+1,k-1}\right) +\left(
2q_{i,j+1,k}-q_{ijk}-q_{i,j+2,k}\right)  \nonumber \\
&&+2\left( 2q_{i,j,k-1}-q_{i+1,j,k-1}-q_{i-1,j,k-1}\right) +\left(
2q_{i,j,k-1}-q_{ijk}-q_{i,j,k-2}\right)  \nonumber \\
&&\left. +2\left( 2q_{i,j,k+1}-q_{i+1,j,k+1}-q_{i-1,j,k+1}\right) +\left(
2q_{i,j,k+1}-q_{ijk}-q_{i,j,k+2}\right) \right]  \nonumber \\
&&+\frac 1{k_BT}\left[ \left( 2H_{ijk}-H_{i+1,j,k}-H_{i-1,j,k}\right) \right.
\nonumber \\
&&\left. \left. +\left( 2H_{ijk}-H_{i,j+1,k}-H_{i,j-1,k}\right) +\left(
2H_{ijk}-H_{i,j,k+1}-H_{i,j,k-1}\right) \right] \right\}
\end{eqnarray}
This equation is in fact not as complex as it seems to be. Each term in a
bracket is a second-order derivative either of the magnetization or of the
external field, and they will cancel each other if a summation is taken,
guaranteeing the conservation of spin. With lattice constant $a$ we can
transform the above equation (and similar equations for lower dimensions)
into 
\begin{equation}
\frac{\partial q\left( {\bf r},t\right) }{\partial t}=\frac{Da^2}{b+K}\left( 
\frac b{2D}-K\right) \nabla ^2q\left( {\bf r},t\right) -\frac{a^2}{2\left(
b+K\right) }\frac 1{k_BT}\nabla ^2H\left( {\bf r},t\right)
\label{Gaussian-field-q}
\end{equation}
where $D$ is the dimensionality. It reveals an important feature that,
surprisingly, the field and the spins are not coupled, which is mainly a
result of the integration taken from $-\infty $ to $+\infty $. The influence
of the inhomogeneous external field rigorously takes the form of a second
order derivative, with a prefactor getting weaker at higher temperature
(weakened by thermal fluctuations).

{\it Discussions}: (1) If the external field is spatially homogeneous, the
system behavior can be described by a simple diffusion equation\cite
{spin-pair}.

(2) By contrast, if the external field varies solely in space, an
equilibrium state can be obtained by setting $\partial q\left( {\bf r}%
,t\right) /\partial t=0$. We find that $\nabla ^2q\left( {\bf r}\right)
\propto \nabla ^2H\left( {\bf r}\right) $, except at the critical point $%
K_c=b/2D$. The susceptibility, $\chi \sim \nabla ^2q\left( {\bf r}\right)
/\nabla ^2H\left( {\bf r}\right) $, can be obtained. To visualize this
process, we suppose that the shape of the external field is a Gaussian
packet: Above the critical point, the magnetization will reach the stable
equilibrium also in the shape of a Gaussian packet, and $\chi $ is finite
and positive. As the system is cooling and approaching the critical point,
we shall observe that the peak of the magnetization (as well as $\chi $)
tends to positive infinity. With the scaling hypothesis $\chi \sim \left|
T-T_c\right| ^{-\gamma }$, the critical exponent $\gamma =1$ can be (in fact
generally) obtained. When the temperature is below $T_c$, the equilibrium
value of the magnetization (and $\chi $) will suddenly become negative
infinity. As the temperature continues to decrease, the magnetization
becomes flatter, and $\chi $ is finite and negative. In this region the
equilibrium is obviously unstable.

A comparison of the two dynamic versions, Kawasaki's and Glauber's, of the
Gaussian model shall be interesting. In Glauber's version, we can obtain\cite
{Zhu-Yang} 
\begin{equation}
\frac d{dt}\langle \sigma _i(t)\rangle =-\langle \sigma _i(t)\rangle +\frac 1%
b\sum_{\left\langle i,j\right\rangle }K_{i,j}\langle \sigma _j(t)\rangle +%
\frac 1{bk_BT}H_i\left( t\right)
\end{equation}
and 
\begin{equation}
\frac{dM\left( t\right) }{dt}=-\left( 1-\frac{2D}bK\right) M\left( t\right) +%
\frac{H\left( t\right) }{bk_BT},
\end{equation}
where $M=\sum_k\left\langle \sigma _k\right\rangle /N$, and $H=\sum_kH_k/N$.
The external field does not change the critical point either, and we can
also get the equilibrium magnetization by setting $dM/dt=0$. The situation
here is quite similar: Above the critical point, the equilibrium is stable,
and it becomes unstable below the critical point, and $\chi $ changes sign.
In other words, in both versions, the observable state below $T_c$ is a fast
growing dynamic process.

(3) From Eq. (\ref{Gaussian-field-q}) we are able to obtain the whole
dynamic process, and here we discuss two interesting one-dimensional
examples.

{\it The first example:} We assume a time-independent field $H=H_0\cos
\left( kx\right) $, and the initial magnetization is zero. Substitution into
Eq. (\ref{Gaussian-field-q}) yields a solution, 
\begin{equation}
q\left( x,t\right) =\frac A{{\cal D}}\left( 1-\exp \left( -{\cal D}%
k^2t\right) \right) H_0\cos \left( kx\right) ,
\end{equation}
where ${\cal D}=a^2\left( b/2-K\right) /\left( b+K\right) $, and $%
A=a^2/\left[ 2\left( b+K\right) k_BT\right] $. When $K<K_c=b/2$, ${\cal D}$
is positive, and $q$ will approach the equilibrium value, $q_{eq}=\left( A/%
{\cal D}\right) H_0\cos \left( kx\right) $, with the relaxation time $\tau
=1/{\cal D}k^2$. A typical critical slowing down phenomenon is obvious when $%
K\rightarrow K_c^{-}$, since $\tau \rightarrow \infty $. Interestingly, the
speed, $dq/dt\sim \exp \left( -{\cal D}k^2t\right) $, does not really ''slow
down'' because of the diverging prefactor $A/{\cal D}$, and this is against
our common knowledge about this phenomenon.

{\it The second example}: We assume a time-dependent field $H=H_0\cos \left(
kx-\omega t\right) $, which is a traveling wave, and we obtain a solution of
Eq. (\ref{Gaussian-field-q}) as 
\begin{equation}
q\left( x,t\right) =\frac{Ak^2}{\sqrt{{\cal D}^2k^4+\omega ^2}}H_0\cos
\left( kx-\omega t+\varphi \right) ,\ \varphi =%
\mathop{\rm arccot}%
\left( {\cal D}k^2/\omega \right) .  \label{gaussian-solution}
\end{equation}
It describes a spin wave, which lags in phase with a temperature-dependent
factor $\varphi $. As a counterpart of the hysteresis loop studied
previously, formed by the average magnetization and the spatially uniform
field, here we introduce the concept of {\it local hysteresis loop}, which
is formed by local magnetization and local field. It is a useful tool
characterizing the response to oscillating perturbations in a simple way.
Generally, the local hysteresis loop may vary in space, while in this
specific example it has an elliptical shape that is spatially uniform and
varies with temperature in the way illustrated in Fig. 1.

\subsection{The Ising model}

\label{subsection 2.2}

Now we will turn to study the Ising model. The Hamiltonian is of the same
form as Eq. (\ref{G-Ham}). In the one-dimensional version of Eq. (\ref{ssr-q}%
), we first treat the two combined terms 
\begin{eqnarray*}
&&\sum_{\hat{\sigma}_k,\hat{\sigma}_{k\pm 1}}\hat{\sigma}_kW_{k,k\pm
1}\left( \sigma _k\sigma _{k\pm 1}\rightarrow \hat{\sigma}_k\hat{\sigma}%
_{k\pm 1}\right) \\
&=&\frac{\sigma _ke^{K\left( \sigma _{k\mp 1}\sigma _k+\sigma _{k\pm
1}\sigma _{k\pm 2}\right) +\frac 1{k_BT}\left( H_k\sigma _k+H_{k\pm 1}\sigma
_{k\pm 1}\right) }+\sigma _{k\pm 1}e^{K\left( \sigma _{k\mp 1}\sigma _{k\pm
1}+\sigma _k\sigma _{k\pm 2}\right) +\frac 1{k_BT}\left( H_k\sigma _{k\pm
1}+H_{k\pm 1}\sigma _k\right) }}{e^{K\left( \sigma _{k\mp 1}\sigma _k+\sigma
_{k\pm 1}\sigma _{k\pm 2}\right) +\frac 1{k_BT}\left( H_k\sigma _k+H_{k\pm
1}\sigma _{k\pm 1}\right) }+e^{K\left( \sigma _{k\mp 1}\sigma _{k\pm
1}+\sigma _k\sigma _{k\pm 2}\right) +\frac 1{k_BT}\left( H_k\sigma _{k\pm
1}+H_{k\pm 1}\sigma _k\right) }} \\
&=&\left\{ 
\begin{array}{c}
\frac 12\left( \sigma _k+\sigma _{k\pm 1}\right) ,\ \text{If }\sigma
_k=\sigma _{k\pm 1} \\ 
\frac 12\left( 1-\sigma _k\sigma _{k\pm 1}\right) \tanh \left[ K\left(
\sigma _{k\mp 1}-\sigma _{k\pm 2}\right) +\frac 1{k_BT}\left( H_k-H_{k\pm
1}\right) \right] ,\ \text{If }\sigma _k=-\sigma _{k\pm 1}
\end{array}
\right. .
\end{eqnarray*}
We assume that the variance of the external field between two
nearest-neighboring vertices is small (i.e. when the wavelength is much
longer than the lattice constant), then 
\begin{eqnarray*}
&&\tanh \left[ K\left( \sigma _{k\mp 1}-\sigma _{k\pm 2}\right) +\frac 1{k_BT%
}\left( H_k-H_{k\pm 1}\right) \right] \\
&\approx &\frac 12\left( \sigma _{k\mp 1}-\sigma _{k\pm 2}\right) \tanh 2K+%
\frac 1{k_BT}\left( H_k-H_{k\pm 1}\right) \left[ 1-\frac 12\left( 1-\sigma
_{k\mp 1}\sigma _{k\pm 2}\right) \tanh ^22K\right]
\end{eqnarray*}
Substituting them into Eq. (\ref{ssr-q}), we get the evolving equation of
single spins, 
\begin{eqnarray*}
\frac d{dt}q_k &=&\frac{a^2}2\left[ \left( q_{k+1}-q_k\right) /a-\left(
q_k-q_{k-1}\right) /a\right] /a \\
&&-\frac{3a^2}4\left\{ \left[ \left( q_{k+2}-q_{k-1}\right) /3a-\left(
q_{k+1}-q_{k-2}\right) /3a\right] /a\right\} \tanh 2K \\
&&+\frac{a^2}4\left\{ \left[ \left( \left\langle \sigma _k\sigma
_{k+1}\sigma _{k+2}\right\rangle -\left\langle \sigma _{k-1}\sigma _k\sigma
_{k+1}\right\rangle \right) /a-\left( \left\langle \sigma _{k-1}\sigma
_k\sigma _{k+1}\right\rangle -\left\langle \sigma _k\sigma _{k-1}\sigma
_{k-2}\right\rangle \right) /a\right] /a\right\} \tanh 2K \\
&&-\frac 12\frac 1{k_BT}\left( H_{k+1}-H_k\right) \left[ 1-\left\langle
\sigma _k\sigma _{k+1}\right\rangle -\frac 12\left\langle \left( 1-\sigma
_k\sigma _{k+1}\right) \left( 1-\sigma _{k-1}\sigma _{k+2}\right)
\right\rangle \tanh ^22K\right] \\
&&+\frac 12\frac 1{k_BT}\left( H_k-H_{k-1}\right) \left[ 1-\left\langle
\sigma _k\sigma _{k-1}\right\rangle -\frac 12\left\langle \left( 1-\sigma
_k\sigma _{k-1}\right) \left( 1-\sigma _{k+1}\sigma _{k-2}\right)
\right\rangle \tanh ^22K\right] .
\end{eqnarray*}
If the external field is spatially uniform, we use $\left\langle \sigma
_k^{(3)}\right\rangle $ to denote $\left\langle \sigma _{k-1}\sigma _k\sigma
_{k+1}\right\rangle $, and 
\begin{equation}
\left( \frac{\partial q_k}{\partial t}\right) _M\equiv \frac 12a^2\left( 1-%
\frac 32\tanh 2K\right) \frac{\partial ^2q}{\partial x^2}+\frac 14a^2\tanh 2K%
\frac{\partial ^2\left\langle \sigma ^{(3)}\right\rangle }{\partial x^2}.
\label{nofield-diffusion}
\end{equation}
In Ref.\cite{Kawasaki}, Kawasaki has used local equilibrium approximation
and concluded that 
\begin{equation}
\left( \frac{\partial q}{\partial t}\right) _M={\cal D}\frac{\partial ^2q}{%
\partial x^2},  \label{eq-diffusion}
\end{equation}
where ${\cal D}\propto 1/\chi $ is the diffusion coefficient vanishing near
the critical point. If this is correct in the Ising model ($\tanh 2K_c=1$
for the one-dimensional case), it requires, 
\begin{equation}
\frac{d^2}{dx^2}\left\langle \sigma _k^{(3)}\right\rangle =\frac{d^2q}{dx^2},
\label{relationship}
\end{equation}
to be true at least near the critical point, otherwise the diffusion process
in the Ising model cannot be described by such a simple equation.

In the following we will incorporate the inhomogeneity of the external field
in our discussion. Let the function $f\left( ka\pm a/2\right) $ denote the
whole term 
\[
f\left( ka\pm a/2\right) =1-\left\langle \sigma _k\sigma _{k\pm
1}\right\rangle -\frac 12\left\langle \left( 1-\sigma _k\sigma _{k\pm
1}\right) \left( 1-\sigma _{k\mp 1}\sigma _{k\pm 2}\right) \right\rangle
\tanh ^22K 
\]
and we get 
\begin{equation}
\frac{\partial q}{\partial t}=\left( \frac{\partial q}{\partial t}\right) _M-%
\frac 12\frac{a^2}{k_BT}\left( f\left( x\right) \frac{\partial ^2H}{\partial
x^2}-\frac{\partial H}{\partial x}\frac{\partial f\left( x\right) }{\partial
x}\right) .  \label{field-diffusion}
\end{equation}
Compared with the Gaussian model, we find both similarities and difference.
First, the diffusion term induced by the inhomogeneity of the magnetization
itself retains the same form as Eq. (\ref{nofield-diffusion}), and this is
also true for the Gaussian model. Second, the influence of the external
field is also weakened by a prefactor $1/T$. However, here the field is
coupled to the spins. And, a somewhat surprising prediction of Eq. (\ref
{field-diffusion}) is that the driving force may partly come from the
gradient of the field, which is at the same time coupled to the
inhomogeneity of the spin configuration. For example, there will still be
field-induced ''diffusion'' if the field varies with space linearly and thus 
$\partial ^2H/\partial x^2=0$, provided that the magnetization is
inhomogeneous and thus $\partial f/\partial x\neq 0$, and this case has been
particularly studied in Ref. \cite{driven-ising-1,driven-ising-2}. Beside
this case, a lot of issues will make us even more curious, including, as
studied in the Gaussian model, the effect of a travelling wave: Shall there
also be a similar spin wave with the same time period and a phase lag? Shall
there be new dynamic phases? What will be the shape of the local hysteresis
loop? Here we provide a tentative answer with Monte Carlo (MC) simulations.

In a one-dimensional model of $N$ vertices with periodic boundary condition,
different initial conditions are chosen while the total spin is kept around
zero. We randomly pick a pair of nearest neighboring spins, $\sigma _i$ and $%
\sigma _j$, and exchange them with a probability 
\[
w_{ij}=\min \left\{ 1,\exp \left( -\Delta E_{ij}/k_BT\right) \right\} , 
\]
where $\Delta E_{ij}$ is the change in energy if $\sigma _i$ and $\sigma _j$
are exchanged. A MC step consists of $N$ such actions. Time is measured in
MC steps and the system is assumed to be of unity length. A travelling wave, 
$H=H_0\cos \left( kx-\omega t\right) $ is applied.

Fig. 2 shows the local hysteresis loops of the $1$st spin in a $100$-spin
chain at several typical temperatures, with $\omega =2\pi /1000$, $k=2\pi $
and $H_0=4$ (in units of $J=k_B$). (The shape of a loop is found to be
independent of the position of the spin it describes.) Each point on a loop
actually denotes the expected value of the spin, i.e., $p_{+}-p_{-}$, where $%
p_{+}$ is the probability that the spin takes upward direction and $p_{-}$
the contrary. The results are to a great degree similar to the rigorous ones
obtained in the Gaussian model. At $K=0.001$ (Fig. 2(a)), the spin
fluctuates around zero (actually it may be a very narrow ellipse with a tiny
phase lag). At $K=0.05$ (Fig. 2(b)), the loop is a tilted ellipse with a
finite phase lag between $0$ and $\pi /2$. At about $K=0.4\sim 0.6$ ($K=0.5$
in Fig. 2(c)), the loop evolves into a state that has two symmetries. It is
symmetric about the horizontal axis, and the average magnetization over a
period is zero. At the same time it is also symmetric about the vertical
axis, and this indicates a phase lag of approximately $\pi /2$. In the
Gaussian model it agrees with the critical point, and here it also
corresponds to a region where a dynamic symmetry loss occurs. First, the
loss of the vertical symmetry: as temperature continues to decrease, the
loops begin to transform into a shape which we are familiar with ($K=2.5$ in
Fig. 2(d)). It means that, though the spin value still varies with the same
period, it no longer varies sinusoidally. Second, the loss of the horizontal
symmetry: for a given spin, the average value over a period may increasingly
deviate from zero as temperature decreases, while the extent and direction
of the deviation are influenced by fluctuations and initial conditions. This
may be accepted as an inherent property of the one-dimensional Ising model
and is observed in the quasi-static limit. We find that the temperature at
which a noticeable symmetry loss occurs is lowered by the magnetic wave.

Here, we tentatively reveal the interesting features that result from a
competition of the external and the internal inhomogeneities and time
scales. A thorough study is needed to answer questions such as: What is the
effect of wave speed, frequency and amplitude? What is the relationship
between the area of the local hysteresis loop and the parameters (previous
works in non-conserved dynamics have revealed several interesting scaling
laws)? In higher dimensions, the effect of a magnetic wave on the formation
and destruction of domains also deserves further investigation. In the case
of a plane wave, we may expect to observe two coexisting length scales, one
parallel to the wave direction, and the other perpendicular to it. This
requires extensive MC simulations which are beyond the scope of this article.

\section{Inhomogeneity of the temperature}

\label{Section 3}

In Sec. \ref{Section 2}, we have investigated the field-induced diffusion.
As mentioned in the Introduction, the inhomogeneity, as the driving force of
the diffusion, may be actually of a broad meaning. In this section we shall
focus on the influence of inhomogeneous temperature. Inspired by the dynamic
response to a periodically altered temperature experimentally observed in
ferroelectric systems\cite{the-hys-fe} and charge-density-wave systems\cite
{the-hys-cdw}, there has been theoretical effort in simple spin models\cite
{the-hys-1,the-hys-2}. With spin-nonconserved dynamics, interesting behavior
and thermal hysteresis have been reported. In the following we present our
treatment of the Kawasaki dynamics of the Gaussian model and the Ising model.

For simplicity we limit our scope to an one-dimensional model that consists
of $N$ spins. There is no (or homogeneous) external field and each vertex, $%
\sigma _k$, is in contact with a heat reservoir of temperature $T_k$. Thus
the system's effective Hamiltonian 
\begin{equation}
-\beta {\cal H}_{eff}\left( \left\{ \sigma \right\} \right)
=\sum_{\left\langle i,j\right\rangle }\frac{K_i+K_j}2\sigma _i\sigma _j.
\label{Hamiltonian-temperature}
\end{equation}
where $K_i=J/k_BT_i$. With 
\begin{equation}
\frac{\nabla K}K=-\frac{\nabla T}T  \label{differential1}
\end{equation}
and 
\begin{equation}
\frac{\nabla ^2K}K=-\frac{\nabla ^2T}T+2\left( \frac{\nabla T}T\right) ^2,
\label{differential2}
\end{equation}
we shall treat $K$ for convenience in the following.

\subsection{The Gaussian model}

\label{subsection3.2}

In the Gaussian model, we substitute the new Hamiltonian, Eq. (\ref
{Hamiltonian-temperature}), into the evolving equation of single spin, Eq. (%
\ref{ssr-q}), and get 
\begin{eqnarray}
\frac d{dt}q_k\left( t\right) &=&-2q_k+\frac 1{2\left[ b+\left(
K_k+K_{k+1}\right) /2\right] }\left[ \left( \frac{K_k+K_{k-1}}2\right)
q_{k-1}\right.  \nonumber \\
&&\left. +\left( \frac{K_k+K_{k+1}}2\right) \left( q_k+q_{k+1}\right)
-\left( \frac{K_{k+1}+K_{k+2}}2\right) q_{k+2}+b\left( q_k+q_{k+1}\right)
\right]  \nonumber \\
&&+\frac 1{2\left[ b+\left( K_k+K_{k-1}\right) /2\right] }\left[ \left( 
\frac{K_k+K_{k+1}}2\right) q_{k+1}\right.  \nonumber \\
&&\left. +\left( \frac{K_k+K_{k-1}}2\right) \left( q_k+q_{k-1}\right)
-\left( \frac{K_{k-1}+K_{k-2}}2\right) q_{k-2}+b\left( q_k+q_{k-1}\right)
\right] .  \label{q-K}
\end{eqnarray}
Substituting 
\begin{equation}
K_{k\pm 1}\approx K_k\pm a\frac{dK}{dx}+\frac 12a^2\frac{d^2K}{dx^2},K_{k\pm
2}\approx K_k\pm 2a\frac{dK}{dx}+2a^2\frac{d^2K}{dx^2},  \label{1}
\end{equation}
and 
\begin{equation}
q_{k\pm 1}\approx q_k\pm a\frac{dq}{dx}+\frac 12a^2\frac{d^2q}{dx^2},q_{k\pm
2}\approx q_k\pm 2a\frac{dq}{dx}+2a^2\frac{d^2q}{dx^2},  \label{2}
\end{equation}
into Eq. (\ref{q-K}), we get 
\begin{eqnarray}
\frac 1q\frac{\partial q}{\partial t} &=&a^2\frac{3b/2-K^{\prime }}{%
K^{\prime }}\left( \frac 1q\frac{\partial ^2q}{\partial x^2}\right) -\frac{%
a^2\left( 2K^{\prime }+3b\right) }{2K^{\prime }}\left( \frac 1q\frac{%
\partial q}{\partial x}\right) \left( \frac 1{K^{\prime }}\frac{dK^{\prime }%
}{dx}\right)  \nonumber \\
&&+a^2\left( \frac 1{K^{\prime }}\frac{dK^{\prime }}{dx}\right) ^2-a^2\left( 
\frac 1{K^{\prime }}\frac{d^2K^{\prime }}{dx^2}\right) ,  \label{Gaussian-K}
\end{eqnarray}
where $K^{\prime }=K+b$. The first term on the right hand side denotes the
diffusion induced by the inhomogeneity of the magnetization itself, which we
are familiar with. The influence of the inhomogeneity of $K$ is described by
the other three terms. The temperature and spin are partially coupled. It is
well established that the magnetization--induced diffusion will vanish near
the critical point (in the one-dimensional model $K_c=b/2$), but {\it the }$%
K ${\it -induced diffusion will not} ({\it neither will the field-induced one%
}). With Eqs. (\ref{differential1}) and (\ref{differential2}), we can see
that the role played by the thermal noise is more difficult to analyze in $K$%
-induced diffusion, compared with the case of field-induced diffusion.
Besides, the third term on the right hand side is always positive, yet,
surprisingly, this is still not against the conservation of the
order-parameter. We can use a simple example to show it. Assume that at one
moment the magnetization is homogeneous, and we have 
\[
\int_{-\infty }^{+\infty }\frac{\partial q}{\partial t}dx=a^2q\int_{-\infty
}^{+\infty }\left[ \left( \frac 1{K^{\prime }}\frac{dK^{\prime }}{dx}\right)
^2-\left( \frac 1{K^{\prime }}\frac{d^2K^{\prime }}{dx^2}\right) \right]
dx=-a^2q\int_{-\infty }^{+\infty }d\left( \frac 1{K^{\prime }}\frac{%
dK^{\prime }}{dx}\right) =0. 
\]

The previous studies cited above mainly considered non-conserved dynamics
when the temperature is varying with time but spatially homogeneous.
Although a similar study of local thermal hysteresis is possible, here we
turn to treat the contrary situation: spatially modulated but temporally
fixed temperature. In the following we visualize the equilibrium state (if
there is one) and the evolving process in three typical examples. The
numerical simulations are based on Eq. (\ref{q-K}), and the system consists
of $100$ spins with periodic boundary condition. We set $b=1$ and assume
that when $t=0$, all the spins take the value of unity.

{\it The first example}: The overall system is above the critical point ($%
\bar{K}<K_c=1/2$) and $K_i=1/4+\left( 1/8\right) \sin \left( 2i\pi
/100\right) $, $i=1,2,...,100$. The evolving process is shown in Fig. 3. We
can clearly see that the system is approaching an equilibrium state and is
slowing down as time passes. Changing the parameters, but keeping $q_i\left(
0\right) =1$ and $K_{\max }<K_c$, we find that, interestingly, the spins
keep to be positive in the evolution.

{\it The second example:} The whole system is below the critical point and $%
K_i=1+\left( 1/8\right) \sin \left( 2i\pi /100\right) $, $i=1,2,...,100$.
Here we may expect the system to be fast growing. The results show an
interesting self-organizing process, which can be characterized by two
succeeding dynamic phases, a steady one, and a growing one. Fig. 4(a) shows
a typical spin configuration during the period between $t=0$ and
approximately $t=400$. The magnetization takes a sinusoidal-like shape and
the amplitude is growing very slowly. After the amplitude reaches a
threshold value, we can observe the self-organization as shown in Fig. 4(b).
Note that part of the magnetization, $q_{50}$ to $q_{100}$, still seems to
be smooth. In fact, self-organization also occurs in this part, only later
and weaker. After this period, the system moves into a fast growing phase
with a fixed shape, as shown in Fig. 4(c). The contrast in the intensity of
motion exhibited by the two regions, $q_1$ to $q_{49}$ and $q_{50}$ to $%
q_{100}$, persists if we change the temperature to, for example, $%
K_i=2+\left( 1/8\right) \sin \left( 2i\pi /100\right) $. So it may be caused
not by the absolute value of the temperature, but by the mutual influence
between the relatively high temperature region and the relatively low
temperature one.

{\it The third example}: Here we study a rather interesting case, where $%
K_i=1/2+\left( 1/4\right) \sin \left( 2i\pi /100\right) $. We can divide the
system into two regions, $q_1\sim q_{50}$ and $q_{51}\sim q_{100}$. In the
first region, the observed evolution is similar to the second example, but
with less peaks and lower growing speed (it is because the temperature is
higher here). In the second region, however, the system behavior is similar
to the first example, i.e. the magnetization is smooth and approaching
equilibrium. We do not observe any strange phenomenon at $K_{50}=K_c=1/2$.
This is not surprising, and can be predicted with Eq. (\ref{Gaussian-K}).

\subsection{The Ising model}

\label{subsection3.3}

Now we turn to treat the one-dimensional Ising model. Similarly, by
substituting Eq. (\ref{Hamiltonian-temperature}) into Eq. (\ref{ssr-q}), we
get 
\begin{eqnarray}
&&\sum_{\hat{\sigma}_k,\hat{\sigma}_{k+w}}\hat{\sigma}_kW_{k,k\pm 1}\left(
\sigma _k\sigma _{k\pm 1}\rightarrow \hat{\sigma}_k\hat{\sigma}_{k\pm
1}\right)  \nonumber \\
&=&\frac 12\left( \sigma _k+\sigma _{k\pm 1}\right) +\frac 12\left( 1-\sigma
_k\sigma _{k\pm 1}\right) \tanh \left( \frac{K_{\mp 1}+K_k}2\sigma _{k\mp 1}-%
\frac{K_{\pm 1}+K_{k\pm 2}}2\sigma _{k\pm 2}\right)  \nonumber \\
&=&\frac 12\left( \sigma _k+\sigma _{k\pm 1}\right) +\frac 14\left( 1-\sigma
_k\sigma _{k\pm 1}\right) \left( \sigma _{k\mp 1}-\sigma _{k\pm 2}\right)
\tanh \left[ \left( K_{\mp 1}+K_k+K_{\pm 1}+K_{k\pm 2}\right) /2\right] 
\nonumber \\
&&\mp \frac 14\left( 1-\sigma _k\sigma _{k\pm 1}\right) \left( \sigma _{k\mp
1}+\sigma _{k\pm 2}\right) \tanh \left[ \left( K_{\mp 1}+K_k-K_{\pm
1}-K_{k\pm 2}\right) /2\right] .
\end{eqnarray}
Applying Eqs. (\ref{1})-(\ref{2}) and similar equations for $\sigma ^{\left(
3\right) }$, we obtain 
\begin{eqnarray}
\frac{\partial q}{\partial t} &=&\left( \frac{\partial q}{\partial t}\right)
_M+a^2\left[ 1-\tanh ^2\left( 2K\right) \right] \frac{\partial K}{\partial x}%
\left( \frac 12\frac{\partial \sigma ^{\left( 3\right) }}{\partial x}-\frac 3%
2\frac{\partial q}{\partial x}\right)  \nonumber \\
&&+a^2\frac{\partial K}{\partial x}\left( \frac{\partial \sigma ^{\left(
3\right) }}{\partial x}-\frac{\partial q}{\partial x}\right) +a^2\frac{%
\partial ^2K}{\partial x^2}\left( \sigma ^{\left( 3\right) }-q\right) .
\label{Ising-K}
\end{eqnarray}
Compared with the rigorous results of the Gaussian model, there are
similarities and differences. Similar to the Gaussian model, the first term
at the right hand side, $\left( \partial q/\partial t\right) _M$, is the
diffusion induced by the magnetization itself (see Eq. (\ref
{nofield-diffusion})). It will vanish near the zero-temperature critical
point. Besides that we can observe the nonvanishing $K$-diffusion, which
will, in contrast, get infinitely sensitive near this point. A major
difference is that, the terms of Eq. (\ref{Ising-K}) do not take such
relative forms as in Eq. (\ref{Gaussian-K}). Thus, although the influence of 
$dK/dx$ and $d^2K/dx^2$ still depend on the spin configuration, the
dependence is in a different way. The system will get less sensitive when
the temperature is increasing, and we may attribute this to the increasing
influence of the thermal noise.

\section{Summary}

\label{Section4}

The present research is based on the Kawasaki-type spin-pair redistribution
mechanism. We generalize the idea of spin diffusion incorporating a variety
of factors as possible driving forces, including external field and
temperature. Two models are selected for a detailed investigation: the Ising
model, which is widely applicable, and the Gaussian model, which is
idealized but provides a good basis for an analytical treatment. It is found
that the two models are similar in principle, and the features are likely to
be shared by other models and experimental systems.

Generally speaking, the diffusion equation can be written as 
\[
\frac{\partial q}{\partial t}=\left( \frac{\partial q}{\partial t}\right)
_M+\left( \frac{\partial q}{\partial t}\right) _{env}. 
\]
The right-hand side of the equation consists of two juxtaposed terms: One
describes the diffusion induced by the inhomogeneity of the magnetization
itself. It retains the same form as that obtained without any external
inhomogeneities, and is believed to be vanishing near the critical point.
The other one is the ''general diffusion'' induced by the inhomogeneity of
the environment, which may be coupled to the spin and contain both first and
second derivatives. The latter, of which a more appropriate name is
environment-induced self-organization, generally does not vanish near the
critical point, and strongly depends on spin configuration and may be
weakened by thermal noise. The concept of local hysteresis is proposed in
this spin-conserved dynamics as a convenient tool to characterize the
response of the system to oscillating external perturbations. The response
of both the Gaussian model and the one-dimensional Ising model to a
travelling electromagnetic wave is specifically studied.

About the Kawasaki dynamics with external inhomogeneities, there remain a
number of questions worthy of further investigations. For example, the
domain growth in an Ising model subject to a travelling electro-magnetic wave%
\footnote{%
Such processes are subject to several factors acting together: the inherent
properties of the model, the moduling effect of the field, and a competion
of time scales, namely the relaxation time of the system and the time period
of the local field.}, the response to the temperature varying with time, and
the dynamics in other models with continuous symmetry, such as the $XY$ model%
\cite{xy} and the Heisenberg model, etc. We hope the interesting features
revealed in this article may stimulate future research, probably with
extensive numerical simulations.

\acknowledgments

This work was supported by the National Natural Science Foundation of China
under Grant No. 10075025.

\appendix 

\section{The spin-pair redistribution mechanism}

\label{Appendix}

In spin-pair redistribution mechanism\cite{spin-pair,compete}, two
neighboring spins, $\sigma _j\sigma _l$, may change to any possible values, $%
\hat{\sigma}_j\hat{\sigma}_l$, as long as their sum are conserved. The
master equation is 
\begin{eqnarray}
\frac d{dt}P(\{\sigma \},t) &=&\sum_{\left\langle jl\right\rangle }\sum_{%
\hat{\sigma}_j,\hat{\sigma}_l}\left[ -W_{jl}\left( \sigma _j\sigma
_l\rightarrow \hat{\sigma}_j\hat{\sigma}_l\right) P\left( \left\{ \sigma
\right\} ;t\right) \right.  \nonumber \\
&&\left. +W_{jl}\left( \hat{\sigma}_j\hat{\sigma}_l\rightarrow \sigma
_j\sigma _l\right) P\left( \left\{ \sigma _{i\neq j},\sigma _{l\neq
k}\right\} ,\hat{\sigma}_j,\hat{\sigma}_l;t\right) \right] .
\label{ssr-master}
\end{eqnarray}
The redistribution probability is in a normalized form determined by a heat
Boltzmann factor, 
\begin{equation}
W_{jl}\left( \sigma _j\sigma _l\rightarrow \hat{\sigma}_j\hat{\sigma}%
_l\right) =\frac 1{Q_{jl}}\delta _{\sigma _j+\sigma _l,\hat{\sigma}_j+\hat{%
\sigma}_l}\exp \left[ -\beta {\cal H}_{jl}\left( \hat{\sigma}_j,\hat{\sigma}%
_l,\left\{ \sigma _m\right\} _{m\neq j,l}\right) \right] ,  \label{ssr-w}
\end{equation}
where the normalization factor $Q_{jl}$ is 
\[
Q_{jl}=\sum_{\hat{\sigma}_j,\hat{\sigma}_l}\delta _{\sigma _j+\sigma _l,\hat{%
\sigma}_j+\hat{\sigma}_l}\exp \left[ -\beta {\cal H}_{jl}\left( \hat{\sigma}%
_j,\hat{\sigma}_l,\left\{ \sigma _m\right\} _{m\neq j,l}\right) \right] . 
\]
For single spins, the time expectation, $q_k\equiv \left\langle \sigma
_k\right\rangle \equiv \sum_{\left\{ \sigma \right\} }\sigma _kP\left(
\left\{ \sigma \right\} ;t\right) $, is 
\begin{equation}
\frac d{dt}q_k\left( t\right) =-2Dq_k\left( t\right) +\sum_w\sum_{\left\{
\sigma \right\} }\left[ \sum_{\hat{\sigma}_k,\hat{\sigma}_{k+w}}\hat{\sigma}%
_kW_{k,k+w}\left( \sigma _k\sigma _{k+w}\rightarrow \hat{\sigma}_k\hat{\sigma%
}_{k+w}\right) \right] P\left( \left\{ \sigma \right\} ;t\right) ,
\label{ssr-q}
\end{equation}
where $D$ is the spatial dimensionality and $\sum_w$ denotes a summation
taken over the nearest neighbors.

\null\vskip0.2cm

\centerline{\bf Caption of figures} \vskip1cm

Fig. 1. Schematic plots of typical local hysteresis loops in the Gaussian
model.

Fig. 2. Typical local hysteresis loops for different temperatures in the
Ising model. System size $N=100$, and parameters $\omega =2\pi /1000$, $%
k=2\pi $ and $H_0=4$.

Fig. 3. The evolution of the system above the critical point. The
magnetization is approaching the equilibrium with its shape similar to the
external field. Squares correspond to the system being at $t=50s$, circles
correspond to $t=300s$, and triangles correspond to $t=1000s$.

Fig. 4. The evolution of the system below the critical point. (a) The
magnetization at $t=350s$ is in the steady phase and keeps a sinusoidal
shape while growing very slowly. (b) The self-organization begins, and the
magnetization at $t=400s$ is shown. (c) The system is in the growing phase
and the magnetization is fast growing while keeping an fixed shape. Squares
correspond to $t=500s$, circles correspond to $t=510s$, and triangles
correspond to $t=520s$.

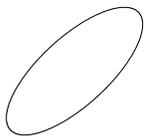
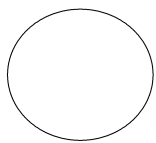
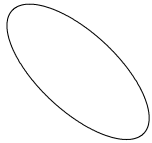
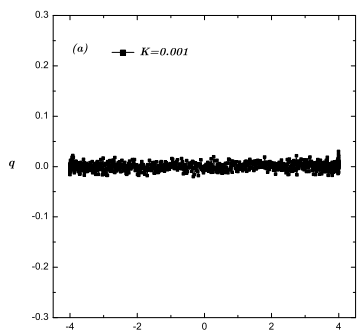
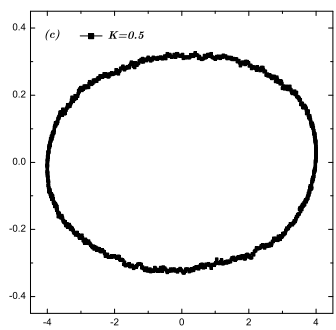
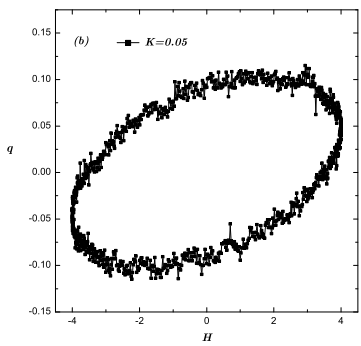
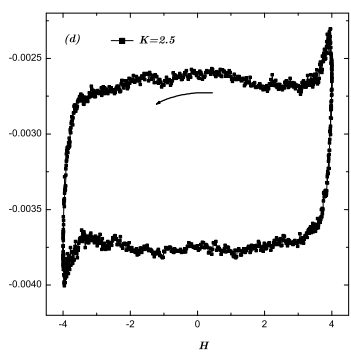
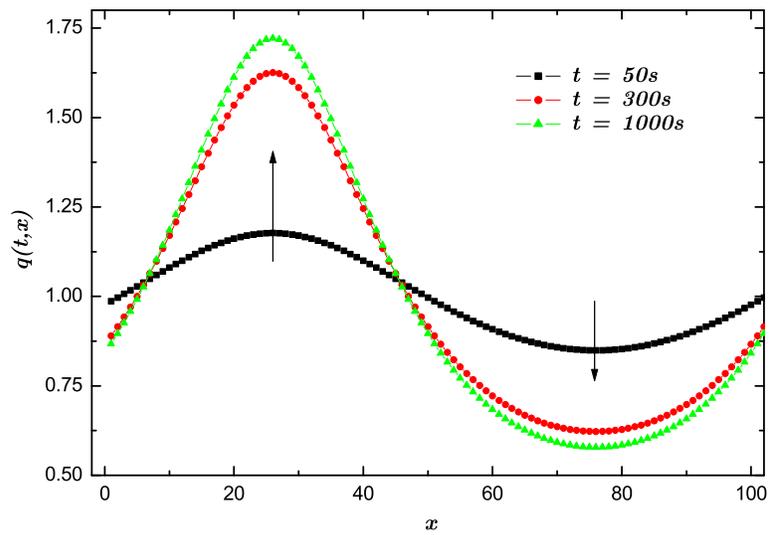
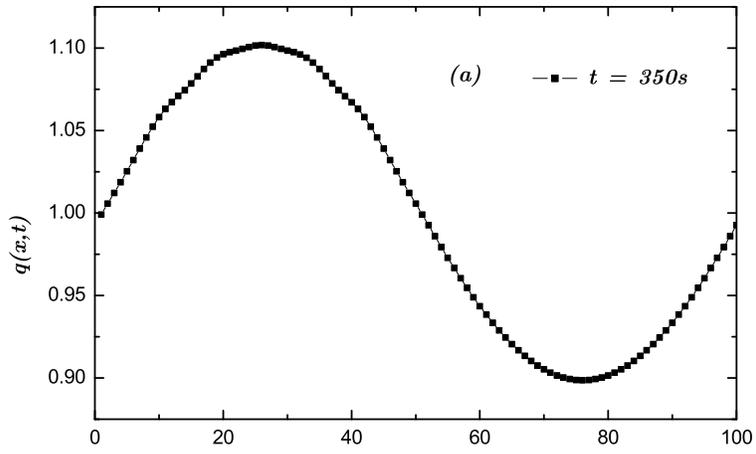
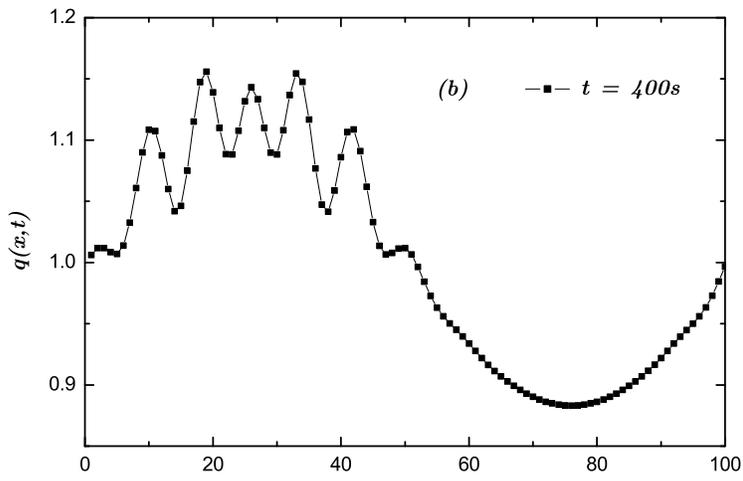
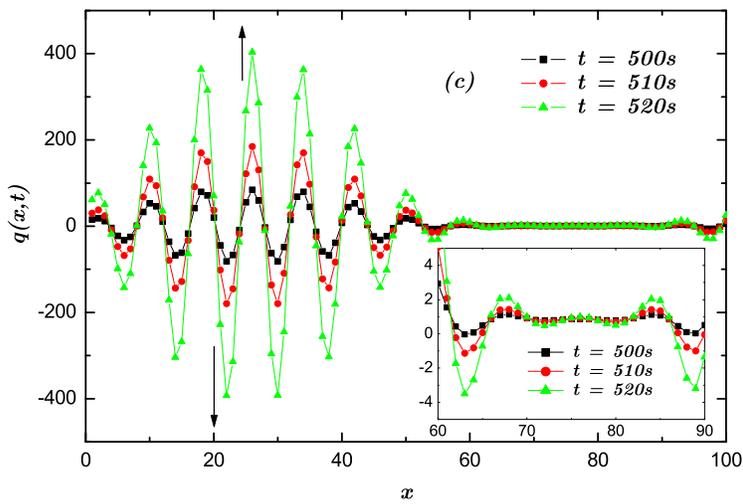
\end{document}